\documentclass[aps,prl,twocolumn,groupedaddress]{revtex4}

\usepackage{graphicx}% Include figure files
\usepackage{dcolumn}% Align table columns on decimal point
\usepackage{bm}% bold math

\begin{document}

%\title{High field ordered phase and upper critical field of the
%filled skutterudite system Pr$_{1-x}$Nd$_x$Os$_4$Sb$_{12}$}
\title{Superconductivity, magnetic order, and quadrupolar order in the filled skutterudite system Pr$_{1-x}$Nd$_{x}$Os$_4$Sb$_{12}$}

\author{P.-C. Ho$^{1,2}, $T.~Yanagisawa$^{1\dagger}$, W.~M.~Yuhasz$^{1\ddagger}$, A.~A. Dooraghi$^1$,
C.~C. Robinson$^1$, N.~P. Butch$^{1\ast}$, R. E.
Baumbach$^1$, and M.~B.~Maple$^1$,} \affiliation {1.~Department of Physics, University of
California, San Diego, La Jolla, CA 92093-0319, U.S.A.\\2. Department of Physics, California
State University, Fresno, CA 93740-8031, U. S. A.}
\altaffiliation {
$\dagger$~Currently at Creative Research Initiative `Sousei', Hokkaido University, Sapporo 001-0021, Japan\\
$\ddagger$~Currently at Ames Laboratory, Ames, IA 50011, U.S.A.\\
$\ast$~Currently at University of Maryland, U.S.A.}

%updated on 4/27/2010, Version 4B

\begin{abstract}
Superconductivity, magnetic order, and quadrupolar order have been investigated in the filled skutterudite system Pr$_{1-x}$Nd$_{x}$Os$_4$Sb$_{12}$ as a function of composition $x$ in magnetic fields up to 9 tesla and at temperatures between 50 mK and 10 K. Electrical resistivity measurements indicate that the high field ordered phase (HFOP), which has been identified with antiferroquadruoplar order, persists to $x$ $\sim$ 0.5. The superconducting critical temperature $T_c$ of PrOs$_4$Sb$_{12}$ is depressed linearly with Nd concentration to $x$ $\sim$ 0.55, whereas the Curie temperature $T_{FM}$ of NdOs$_4$Sb$_{12}$ is depressed linearly with Pr composition to ($1-x$) $\sim$ 0.45. In the superconducting region, the upper critical field $H_{c2}(x,0)$ is depressed quadratically with $x$ in the range 0 $<$ $x$ $\lesssim$ 0.3, exhibits a kink at $x$ $\approx$ 0.3, and then decreases linearly with $x$ in the range 0.3 $\lesssim$ $x$ $\lesssim$ 0.6. The behavior of $H_{c2}(x,0)$ appears to be due to pair breaking caused by the applied magnetic field and the exhange field associated with the polarization of the Nd magnetic moments, in the superconducting state. From magnetic susceptibility measurements, the correlations between the Nd moments in the superconducting state appear to change from ferromagnetic in the range 0.3 $\lesssim$ $x$ $\lesssim$ 0.6 to antiferromagnetic in the range 0 $<$ $x$ $\lesssim$ 0.3. Specific heat measurements on a sample with $x$ $=$ 0.45 indicate that magnetic order occurs in the superconducting state, as is also inferred from the depression of $H_{c2}(x,0)$ with $x$.

\end{abstract}

% insert suggested PACS numbers in braces on next line
% insert suggested keywords - APS authors don't need to do this

%\maketitle must follow title, authors, abstract, \pacs, and \keywords
\maketitle

% body of paper here - Use proper section commands
% References should be done using the \cite, \ref, and \label commands
% Put \label in argument of \section for cross-referencing
\section{introduction}
Since the discovery of heavy fermion (HF) superconductivity (SC) in PrOs$_4$Sb$_{12}$ in 2001,~\cite{bauer_2002_1} this compound has attracted intense interest. It is one of the few Pr compounds that display HF behavior (other examples include PrInAg$_2$~\cite{yatskar_1996_1} and PrFe$_4$P$_{12}$~\cite{aoki_2002_1}), it has a high field ordered phase (HFOP), for magnetic fields between 4.5 T and 14.5 T and temperatures below 1 K, ~\cite{Ho2003,vollmer2003,Oeschler2004,Aoki2002,Tayama2003,Rotundu2004} that has been identified with antiferroquadrupolar order,~\cite{Kohgi2003} and it exhibits some type of unconventional superconductivity.~\cite{maple_2006_1,Aoki2005} The occurence of the HFOP can be traced to the small splitting between the ground and first excited states of the Pr$^{3+}$ Hund's rule multiplet by the crystalline electric field (CEF).~\cite{maple_2006_1,Aoki2005} The CEF
energy level scheme has been studied in detail, and is consistent with a non-magnetic $\Gamma_1$ singlet
ground state (0\,K), a low-lying magnetic $\Gamma_4^{(2)}$ triplet
first excited state (\mbox{$\sim$ 7\,K}), a $\Gamma_4^{(1)}$
magnetic triplet excited state (\mbox{$\sim$ 130\,K}), and a
non-magnetic $\Gamma_{23}$ doublet excited state (\mbox{$\sim$
200\,K}), in T$_{\rm{h}}$
symmetry.~\cite{Goremychkin2004} Although a
$\Gamma_1$ ground state does not typically give rise to HF
behavior, an intriguing possibility is that the HF state and unconventional superconductivity originate from electric quadrupole fluctuations, in the vicinity of a quadrupolar quantum critical point.  Various measurements have also provided evidence for multiple SC phases,
two-band SC, and point nodes in the SC energy gap.~\cite{Izawa2003,Grube2006,Seyfarth2005,Chia2003} Moreover, an internal magnetic field was detected in the SC
state by $\mu$SR measurements,~\cite{Aoki2003} indicating an unconventional SC state and, possibly, spin-triplet pairing mechanism of electrons.

The end member compound NdOs$_4$Sb$_{12}$ is a mean-field type ferromagnet with a Curie temperature $T_{FM}$ $\approx$ 1 K. According
to the analysis of magnetic susceptibility, electrical resistivity,
and ultrasonic attenuation data, the CEF energy level scheme of the Nd$^{3+}$
ion in NdOs$_4$Sb$_{12}$ is consistent with a $\Gamma_8^{(2)}$ quartet
ground state (0\,K),  a $\Gamma_8^{(1)}$ quartet first excited state
(\mbox{$\sim$ 220\,K}), and a $\Gamma_6$ doublet highest excited state
(\mbox{$\sim$ 590\,K}), all of which are magnetic in
O$_{\rm{h}}$ symmetry.~\cite{Ho2005,Yanagisawa2007} Recent inelastic
neutron scattering measurements support this energy level scheme
with a more accurate description in T$_{\rm{h}}$ symmetry:
$\Gamma_{67}^{(2)}(0\,K) - \Gamma_{67}^{(1)}(267\,K) - \Gamma_{5}
(350\,K)$.~\cite{Kuwahara2008,Takegahara2001}  Although the CEF ground state $\Gamma_{67}^{(2)}$ contains quadrupole moments, no features
indicative of a HFOP have been detected in NdOs$_4$Sb$_{12}$.  A
large electronic specific heat coefficient $\gamma \sim 520$
\,mJ/(mol$\cdot\rm{K}^2$) is inferred at low $T$.~\cite{Ho2005} In addition to the typical rattling mode that is observed as an ultrasonic dispersion near $\sim 40$\,K in the compounds ROs$_4$Sb$_{12}$ (R= La, Pr, Nd, Sm), an extra rattling mode is detected only in
NdOs$_4$Sb$_{12}$.~\cite{Yanagisawa2008}  Because the FM transition in
NdOs$_4$Sb$_{12}$ occurs at quite a low $T$ and the lattice
parameter is almost the same as that of PrOs$_4$Sb$_{12}$, the
substitution of Pr with Nd in the pseudoternary system
Pr$_{1-x}$Nd$_x$Os$_4$Sb$_{12}$ is well suited to investigating the effect
of FM on the evolution of the unconventional SC of
PrOs$_4$Sb$_{12}$.

In this paper, we report the temperature vs Nd concentration $T-x$ phase diagram of the Pr$_{1-x}$Nd$_x$Os$_4$Sb$_{12}$ system. We find that the superconducting critical temperature $T_c$ of the Pr end member compound decreases linearly with Nd concentration $x$, while the Curie temperature of the ferromagnetic end member compound decreases linearly with the Pr concentration ($1-x$). The superconducting region extends to $x$ $\approx$ 0.55, where it meets the ferromagnetic phase, with evidence for ferromagnetic ordering of the Nd moments in the superconducting phase at $x$ $\approx$ 0.45, from specific heat measurements in magnetic fields up to 0.429 tesla. In order to gain information about the evolution of the HFOP with Nd concentration and to probe the nature of the superconducting state, magnetoresistive measurements were performed in fields up to 9 T throughout the range 0 $\lesssim$ $x$ $\lesssim$ 0.5. An analysis of the upper critical field $H_{c2}(x,0)$, involving a comparison to measurements of $H_{c2}(x,0)$ on the La$_{3-x}$Gd$_x$In system and the multiple pair breaking theory of Fulde and Maki, provides evidence for magnetic ordering of the Nd ions in the superconducting state for concentrations above $x$ $\approx$ 0.3 and suggests that the superconducting electrons of Pr$_{1-x}$Nd$_x$Os$_4$Sb$_{12}$ are paired in spin singlet states. On the other hand, the evolution of $H_{c2}(x,0)$ might be described in a two band model of superconductivity. Magnetic susceptibility measurements indicate that the magnetic correlations of the Nd ions change from ferromagnetic for 0.3 $\lesssim$ $x$ $\lesssim$ 0.6 to antiferromagnetic for 0 $<$ $x$ $\lesssim$ 0.3.

\section{Experimental Details}
\begin{figure}
 \begin{center}
  \includegraphics[width=3.35in]{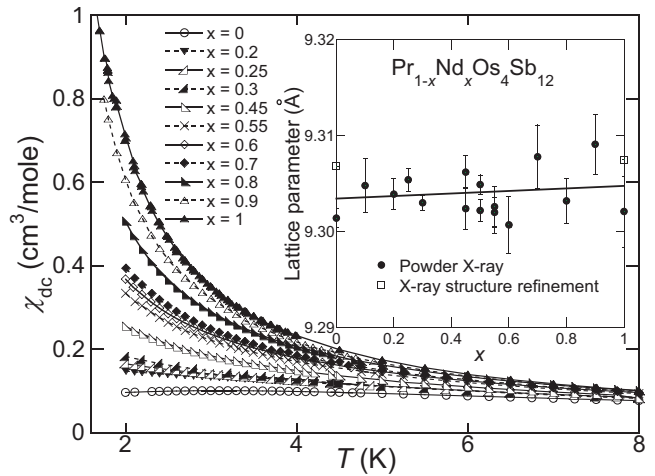}
   \caption{DC molar magnetic susceptibility $\chi_{dc}$ as a function of temperature $T$ for 2\,K - 8\,K for single crystals of Pr$_{1-x}$Nd$_x$Os$_4$Sb$_{12}$
            at various Nd concentrations $x$.
            Inset: Nd concentration $x$ dependence of the lattice parameter.}
  \label{fig:chivsTLattvsNdx}
 \end{center}
\end{figure}

Single crystals of Pr$_{1-x}$Nd$_x$Os$_4$Sb$_{12}$ were grown by the
molten flux method as described in Ref~\onlinecite{BauerED2001}. Pr
and Nd were mixed in an arc furnace prior to the flux growth to
ensure a uniform distribution of the rare earth elements. The cubic
LaFe$_4$P$_{12}$-type structure~\cite{Jeitschko1977} is observed by means of X-ray powder diffraction
measurements
throughout the entire doping series. The lattice parameter vs Nd concentration $x$ is
plotted in the inset of Fig.~\ref{fig:chivsTLattvsNdx}, where it is
apparent that the lattice constant is minimally affected by Nd
substitution, consistent with previous
measurements by Jeitschko et al.~\cite{Jeitschko1977}
%(average size $\sim 9.304$\,$\rm{\AA}$
% fig.1: ChixvsTLattvsx_PrNdOsSb.eps
The DC magnetic susceptibilities
$\chi_{\rm{dc}}(T)$ (Fig.~\ref{fig:chivsTLattvsNdx}) for collections of single crystals with a total mass
near 30\,mg were measured using a Quantum Design
SQUID magnetometer MPMS-5.5. Together with the X-ray diffraction data, the $\chi_{\rm{dc}}(T)$ data show that Nd can be substituted continuously for all values of $x$: i.e., higher
values of $\chi_{\rm{dc}}(T)$ are observed in samples with higher
$x$. The ac magnetic susceptibility $\chi_{\rm{ac}}(T)$ measurements were performed using home-built 1st-order gradiometers as pick-up coils in the temperature
range from 0.05 K to 2.5 K.  Each pick-up coil is coupled with a primary coil,
which supplies a 17 Hz $\sim 0.05-0.15$\,Oe ac magnetic field. Typically, $\sim 6 -
18$\, mg collections of single crystals were used
for the $\chi_{\rm{ac}}(T)$ measurements. Electrical resistivity measurements $\rho(T, H)$ were performed using a standard 4-wire technique in a transverse geometry ($H \perp$ current) on individual single crystals mounted in a $^3$He-$^4$He dilution refrigerator in magnetic fields $H$ between 0\,T and 9\,T. Specific heat $C(T,H)$ measurements were performed on a collection of single crystals with a mass of 51.07 mg for $x$ $=$ 0.45 using a standard heat pulse technique.

\section{Results}

\begin{figure}
 \begin{center}
   \includegraphics[width=3.35in]{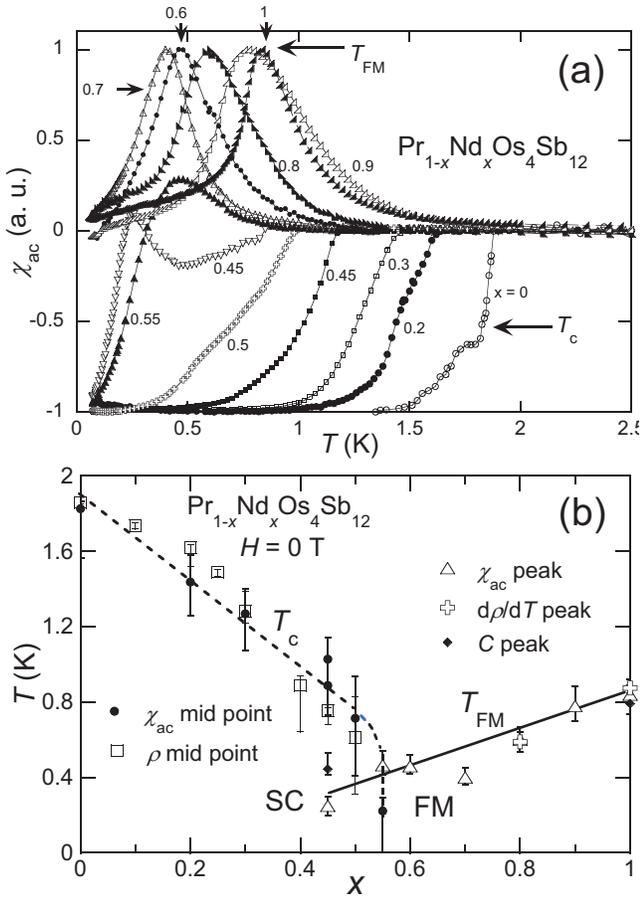}
   \caption{(a) Temperature $T$ dependence of the ac magnetic susceptibility $\chi_{ac}$
                in arbitrary units for various Nd concentrations $x$.
            (b) Superconducting transition temperature $T_{\rm{c}}$ and ferromagnetic
                transition (Curie) temperature $T_{\rm{FM}}$ vs $x$ for Pr$            _{1-x}$Nd$_x$Os$_4$Sb$_{12}$
                determined from measurements of the ac magnetic susceptibility $\chi_{\rm{ac}}$,
                electrical resistivity $\rho$, and specific heat $C$.
                Vertical bars indicate transition widths as defined in the text.}
  \label{fig:TcTFMvsNdx}
 \end{center}
\end{figure}

Displayed in Fig.~\ref{fig:TcTFMvsNdx}(a) are the ac magnetic
susceptibility data $\chi_{ac}(T)$ for various $x$, where the paramagnetic background signal at 2.5\,K has been set to
zero and the data for each $x$ have been normalized to the largest signal
in either the superconducting (SC) or the ferromagnetic (FM)
transitions.  When a sample enters the SC state, $\chi_{ac}(T)$
drops below the paramagnetic background in a rounded step shape, where
the SC transition temperature $T_{\rm{c}}$ is defined at 50$\%$ of the change in $\chi_{ac}$ and the transition width is defined as the difference in the temperatures associated with the $10\%$ and $90\%$ values. When FM ordering occurs, $\chi_{ac}(T)$ exhibits a
peak above the paramagnetic background at the FM transition (Curie)
temperature $T_{\rm{FM}}$ and the transition width is taken to be the difference in the temperature corresponding to $90 \%$ of the peak value of $\chi_{ac}(T)$ and $T_{FM}$.
As $x$ increases, the SC transition is
suppressed to lower $T$ (at almost the same rate as for Pr(Os$_{1-x}$Ru$_x$)$_4$Sb$_{12}$~\cite{Ho2008}) and, above $x = 0.6$, only a FM signal
appears. For $x = 0.45$ and 0.55, features associated with both SC and FM
are observed. Figure~\ref{fig:TcTFMvsNdx}(b) summarizes the $x$ dependence of
$T_{\rm{c}}$ and $T_{\rm{FM}}$ determined from measurements of
$\chi_{ac}(T)$, $\rho(T)$, and $C(T)$. We also note
the discrepancy between the two $x = 0.45$ samples in
Fig.~\ref{fig:TcTFMvsNdx}(a), which may be due to supercurrent surface
screening in one of the samples which obscures
the FM feature in the $\chi_{ac}(T)$ measurements. On the other hand, sample
dependence cannot be completely ruled out.

\begin{figure}
\begin{center}
 \includegraphics[width=3.35in]{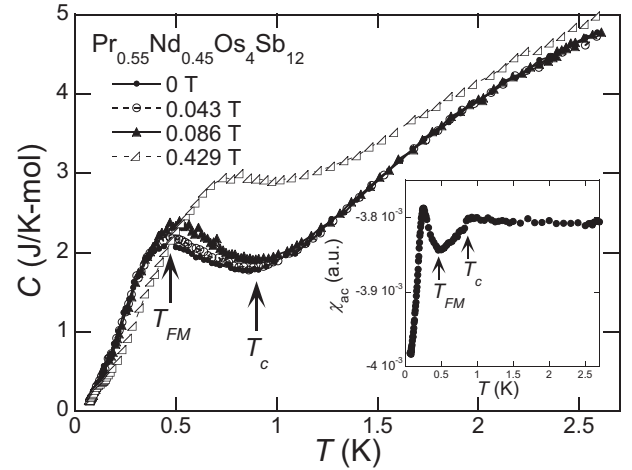}
  \caption{Specific heat $C$ of Pr$_{0.55}$Nd$_{0.45}$Os$_4$Sb$_{12}$ vs
           $T$ at $H$ = 0, 0.043\,T, 0.086T, and 0.429\,T. As $H$ increases,
           the broad peak moves to higher $T$. Inset: the corresponding $\chi_{ac}(T)$
           measured on single crystals from the same batch.}
  \label{fig:CoexistenceSCFMNd0_45}
 \end{center}
\end{figure}
The specific heat of a sample with $x$ $=$ 0.45 was measured in order to explore the possible coexistence of SC and FM at this concentration (Fig.~\ref{fig:CoexistenceSCFMNd0_45}). A broad
peak appears in $C(T,H = 0)$ at 0.9 K with a maximum near 0.48 K.  The inset to
Fig.~\ref{fig:CoexistenceSCFMNd0_45} shows the corresponding
$\chi_{\rm{ac}}(T)$ data, where the onset of $T_{\rm{c}}$ at $\sim
0.9$\,K matches the beginning of the upturn of the peak in $C(T,H = 0)$. The FM feature then appears in $\chi_{\rm{ac}}(T)$ near 0.48 K, in
agreement with the maximum in $C(T, H = 0)$.
From these observations, we conclude that FM and SC features are present in both $\chi_{ac}(T)$ and $C(T, H = 0)$ and that the broadness of the
peak in $C(T, H = 0)$ may be due to the proximity of the two
transitions. Upon application of small magnetic fields, the
peak gradually shifts to higher $T$, indicating that the SC phase is suppressed and the FM phase is enhanced with $H$.

\begin{figure}
 \begin{center}
  \includegraphics[width=3.35in]{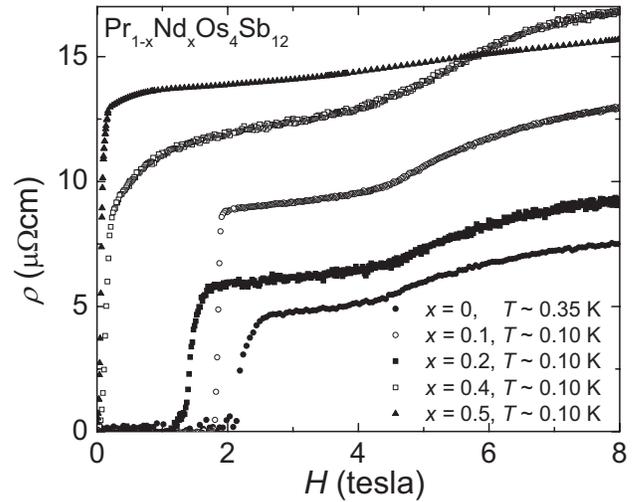}
   \caption{Electrical resistivity $\rho$ vs magnetic field $H$ at various temperatures $T$. The rapid drop in $\rho(T,H)$
to zero is due to the superconducting transition, while
the shoulders in $\rho(H)$ above 4 T are due to the HFOP.~\cite{Ho2003}}
  \label{Fig4}
 \end{center}
\end{figure}
\begin{figure}
 \begin{center}
  \includegraphics[width=3.35in]{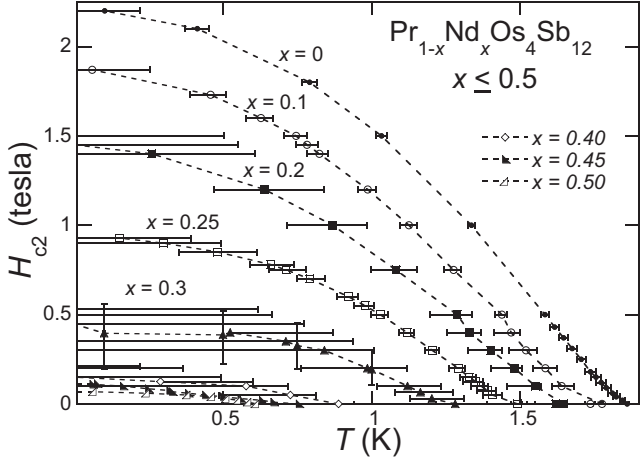}
   \caption{Temperature dependence of upper critical field $H_{\rm{c2}}$
            for Nd concentrations $x \le 0.5$.  The horizontal and vertical bars represent
            the transition widths, defined from the $10\%$ and $90\%$ values of the drop in
            $\rho(T, \rm{fixed~H})$ at $T_{\rm{c}}$ and  $\rho(\rm{fixed~T}, \it{H})$ at
$H_{\rm{c2}}$. Inset: Normalized $H_{\rm{c2}}(x,0)$ and $x$ with respect to
           $H_{\rm{c2}}(0,0)$ and $x_{\rm{cr,1}}$ for
           Pr$_{1-x}$Nd$_x$Os$_4$Sb$_{12}$ and La$_{3-x}$Gd$_x$In.
           Note the difference between the curvature at low concentrations.}
  \label{fig:Hc2vsT}
 \end{center}
\end{figure}

From measurements of $\rho(T)$ at constant $H$ and $\rho(H)$ at fixed $T$ (Fig.~\ref{Fig4}),
$H_{c2}(T)$ curves were determined for various values of $x$ (Fig.~\ref{fig:Hc2vsT}). The
data points are defined as the temperatures and fields associated with the 50$\%$ value of $\Delta \rho$ at the SC
transition, and the transition width is defined as the differences in temperatures and fields corresponding to the 10$\%$ and 90$\%$
values of $\Delta \rho$ at the SC transition.  Above $x =0.25$, the transition width becomes very
large. Measurements of $\rho(H)$ isotherms also reveal the HFOP phase, which appears as a shoulder in $\rho(H)$ above 4 T.~\cite{Ho2003} The $T$ where the HFOP phase appears is defined as the sharp kink where $\rho(H)$ increases with increasing $H$.
% fig.4: Hc2vsT.eps

Since the FM in NdOs$_4$Sb$_{12}$ conforms to the mean field model,
a Curie-Weiss analysis of the Nd contribution to the magnetic susceptibility should be a good indication of
the evolution of FM in the Pr$_{1-x}$Nd$_x$Os$_4$Sb$_{12}$ system.
The contribution to the magnetic susceptibility due to the Pr$^{3+}$
ions is subtracted from $\chi_{\rm{dc}}(T)$ of
Pr$_{1-x}$Nd$_x$Os$_4$Sb$_{12}$ in the following way:
 \begin{equation}
  \chi_{_{\rm{Nd}}}(T) =
  \frac{\chi_{_{\rm{Pr_{1-x}Nd_xOs_4Sb_{12}}}}(T)-(1-x)\chi_{_{\rm{PrOs_4Sb_{12}}}}(T)}{x}.
  \label{eq:chiNdExtraction}
 \end{equation}
We note that in order to apply this type of analysis, it is necessary to assume that the contribution to $\chi_{\rm{dc}}(T)$ from the Pr$^{3+}$ ions must retain the same $T$ dependence for all values of $x$. This seems like a reasonable approximation because the nearest neighbors of the rare earth ions are Sb ions, which form the cages of the filled skutterudite structure. As such, the CEF that influences each Pr ion is, to first order, unchanged as Pr is replaced with Nd. Due to the curvature for 20\,K-50\,K caused by the
effect of the CEF on the Nd$^{3+}$ ions, the
Curie-Weiss analysis of $\chi_{\rm{Nd}}^{-1}(T)$ is only applied in
the low $T$ regime from 2\,K to 10\,K by using the expression:
\begin{equation}
 \chi_{_{\rm{Nd}}} (T)= \frac{\rm{C}}{T - \Theta},
 \label{eq:CW}
\end{equation}
where C is the Curie constant and $\Theta$ is the
Curie-Weiss temperature.  The fitting results are displayed in
Fig.~\ref{fig:CWfit}.  The CW temperature is positive and decreases between $x$ = 1 and $x_{cr,2}$ $\sim$ 0.3, where it crosses over to a negative value. The Curie constant undergoes a modest increase with decreasing $x$, with a possible local maximum near $x_{cr1}$ $\sim$ 0.6.  These results suggest the presence of a weak FM phase between $x \approx 0.3$ and
1 and that AFM correlations appear below $x \approx 0.3$.

\section{Analysis and Discussion}
% fig.5: HvsNdx_PrNdOsSb.eps
Taken together, these results reveal a rich phase diagram for the Pr$_{1-x}$Nd$_x$Os$_4$Sb$_{12}$ system which includes superconductivity, magnetic order, and quadrupolar order.

\begin{figure}
 \begin{center}
  \includegraphics[width=3.35in]{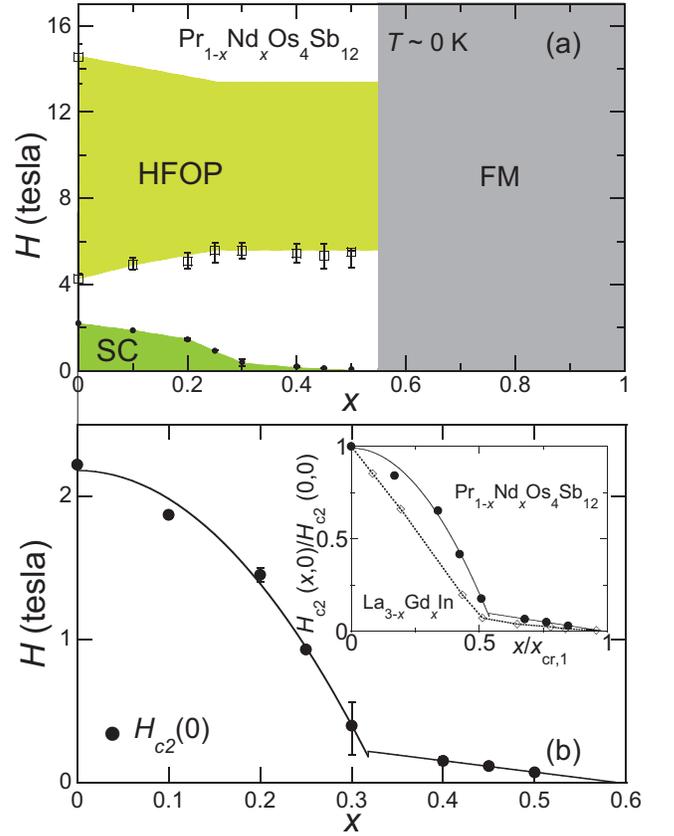}
  \caption{(a) Magnetic field $H-x$ phase diagram of Pr$_{1-x}$Nd$_x$Os$_4$Sb$_{12}$ at $\sim 0$\,K.
           (b) Zero-kelvin extrapolation of the experimentally determined upper critical field H$_{\rm{c2}}$(x,0). The solid line is a curve based on Eqs.~(\ref{eq:Hc20xnumerical}). Inset: Normalized H$_{\rm{c2}}$(x,0) and $x$ with respect to H$_{\rm{c2}}$(0,0) and $x_{cr,1}$ for Pr$_{1-x}$Nd$_x$Os$_4$Sb$_{12}$ and La$_{3-x}$Gd$_x$In. Note the difference between the curvature at low concentration.}
  \label{fig:HvsNdx}
 \end{center}
\end{figure}
Figure~\ref{fig:HvsNdx}(a) shows the 0\,K $H-x$ phase diagram for
Pr$_{1-x}$Nd$_x$Os$_4$Sb$_{12}$.  The lower boundary of the
high field ordered phase (HFOP) is determined from kinks in the
$\rho(H)$ isotherms (Fig.~\ref{Fig4}). The HFOP in Pr$_{1-x}$Nd$_x$Os$_4$Sb$_{12}$
persists to $x=0.5$, above which resistivity measurements have
not yet been performed, although we note that the features associated with the HFOP are not observed for NdOs$_4$Sb$_{12}$. The break in slope in the lower $H-x$ phase boundary of the HFOP seems to be correlated with the one observed in $H_{\rm{c2}}(0) - x$. In addition, the HFOP persists throughout the SC phase in Pr$_{1-x}$Nd$_x$Os$_4$Sb$_{12}$ , whereas it vanishes beyond $x$ $\sim$ 0.1 in
Pr(Os$_{1-x}$Ru$_x$)$_4$Sb$_{12}$.~\cite{Ho2008} This may be due to changes in the CEF that are larger for Ru substitution for Os than for Nd substitution for Pr.

From the $H_{c2}(x,T)$ data shown in Fig.~\ref{fig:Hc2vsT}, it can be seen that the magnitude of $H_{c2}$ decreases rapidly with $x$ in the range 0 $<$ $x$ $\lesssim$ 0.3, more slowly in the range 0.3 $\lesssim$ $x$ $\lesssim$ 0.55, and appears to vanish near $x$ $\approx$ 0.6. This trend is illustrated more clearly in Fig.~\ref{fig:HvsNdx}b, which shows the zero temperature value of the upper critical field $H_{c2}(x,0)$. The $H_{\rm{c2}}(x, 0)$ data can be fit with the equations (solid
line in Figure~\ref{fig:HvsNdx}(b)):
%y = P1 + P2 * x^2, P1 = 2.1798 +- 0.06, P2 = 19.78 +- 1.14, R^2 = 0.99012
\begin{equation}
H_{\rm{c2}}(x,0) \approx
  \left \{
   \begin {array}{ll}
     2.18 - 19.78 x^2 &~0 \le x < 0.3,\\
     0.471 - 0.8 x~~~~~~~~~~~~&~0.3 < x < 0.6.
    \end {array}
      \right\}
 \label{eq:Hc20xnumerical}
\end{equation}
where the $x$ dependence of $H_{\rm{c2}}(0)$ is quadratic for $x \lesssim 0.3$ and is linear for $0.3 \lesssim x \lesssim 0.6$, resulting in an obvious break in slope near $x \sim 0.3$. Thus, there appear to be two critical concentrations, $x_{cr2}$ $\approx$ 0.3 and $x_{cr1}$ $\approx$ 0.6. Since evidence for
two-band SC has been observed in
PrOs$_4$Sb$_{12}$,~\cite{Seyfarth2005}, this result could indicate
that Nd substitution has a different effect on the SC of the different
bands. If so, then this empirical formula may suggest that one
channel has a critical concentration $x_{\rm{cr,1}} \sim 0.6$, for
which $H_{\rm{c2}}$ has a linear $x$ dependence over the entire SC region with
a maximum value of $\sim 0.471$\,T, while the other channel has a critical
concentration $x_{\rm{cr,2}} \sim 0.3$ and only exists in the
region $0 \le x \le 0.3$ with a maximum value of $H_{c2}$ $\sim 2.2$\,T.

% fig.6: CWfitonNd.eps
\begin{figure}
 \begin{center}
  \includegraphics[width=3.35in]{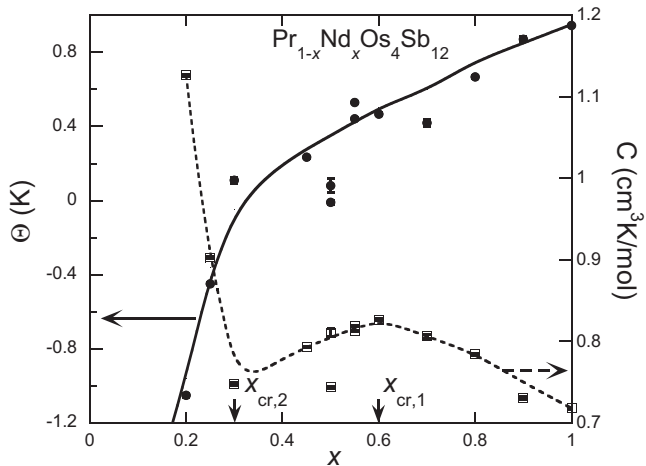}
  \caption{Nd concentration $x$ dependence of the Curie-Weiss temperature
  (left axis) $\Theta$ and the Curie constant C (right axis) of
  the Nd$^{3+}$ ion in Pr$_{1-x}$Nd$_x$Os$_4$Sb$_{12}$ after
  the magnetic susceptibility $\chi_{\rm{dc}}(T)$ of PrOs$_4$Sb$_{12}$
  has been subtracted. The fitting range is between 2\,K and 10\,K. Solid and dashed lines are guides to the eye.}
  \label{fig:CWfit}
 \end{center}
\end{figure}
%low T mu_eff(Pr) =3.09 mu_B, mu_eff(Pr) =2.43 mu_B

% fig.7: NormalizedPairBreakingHc2.eps
%\begin{figure}
% \begin{center}
%  \includegraphics[width=3.35in]{Fig8}
%  \caption{Zero-K extrapolation of the $H_{\rm{c2}}(x,T)$ curves vs Gd
%           concentration $x$ in La$_{3-x}$Gd$_x$In (data from the work of
%           Crow $et$ $al.$ in Ref.~\onlinecite{Crow1967}).
%           The experiments on the La$_{3-x}$Gd$_x$In system addressed the effect of
%           magnetic substituents (Gd$^{3+}$) on conventional superconductivity in La$_3$In.
%           Inset: Normalized $H_{\rm{c2}}(x,0)$ and $x$ with respect to
%           $H_{\rm{c2}}(0,0)$ and $x_{\rm{cr,1}}$ for
%           Pr$_{1-x}$Nd$_x$Os$_4$Sb$_{12}$ and La$_{3-x}$Gd$_x$In.
%           Note the difference between the curvature at low concentrations.}
%  \label{fig:ComparetoLa3_xGdxIn}
% \end{center}
%\end{figure}

In order to explore an alternative route to analyzing the $H_{c2}(x,0)$ data, we turn to the multiple pair breaking theory of Fulde and Maki.~\cite{Fulde1966,Crow1967,Maple1973} For example, this theory has previously been used to analyze the $H_{c2}(x,T)$ curves for the system La$_{3-x}$Gd$_x$In, that is formed by substituting Gd impurity ions that carry localized magnetic moments into the singlet BCS superconductor La$_3$In.~\cite{Crow1967}
The $H_{c2}(x,0)$ data in the La$_{3-x}$Gd$_x$In system reveal that $H_{c2}(x,0)$ exhibits a rapid linear decrease with $x$ in the lower concentration region and a slower linear depression with $x$ in the higher concentration region.
In the inset of Fig.~\ref{fig:HvsNdx}, the normalized upper critical field data $H_{\rm{c2}}(x,0)$/$H_{\rm{c2}}(0,0)$ for both systems are plotted vs the normalized concentration $x/x_{cr1}$. Here, $H_{\rm{c2}}(0,0)$ is the 0 K
$H_{\rm{c2}}$ of the parent compound without magnetic
substituents (i.e., PrOs$_4$Sb$_{12}$ and La$_{3}$In), and $x_{\rm{cr,1}}$ is the concentration where SC disappears. $H_{\rm{c2}}(0) \sim 2.25$\,tesla for PrOs$_4$Sb$_{12}$ and $\sim 6.8$\,tesla for La$_3$In, while $x_{\rm{cr,1}} \sim 0.6$ for Pr$_{1-x}$Nd$_x$Os$_4$Sb$_{12}$ and $\sim 0.076$ for La$_{3-x}$Gd$_x$In. The ratio of $x_{\rm{cr,1}}$ of Pr$_{1-x}$Nd$_x$Os$_4$Sb$_{12}$ to that of La$_{3-x}$Gd$_x$In is
$\sim 7.6$, or $\sim 23$, if the magnetic substituent is
expressed as the percentage of the over all rare-earth concentration in each compound.
For $0 \le x/x_{cr,1} \lesssim 0.5$, the suppression of $H_{\rm{c2}}(x,0)$/$H_{\rm{c2}}(0,0)$ in Pr$_{1-x}$Nd$_x$Os$_4$Sb$_{12}$ is much less than that of La$_{3-x}$Gd$_x$In, but for $0.5 \lesssim x/x_{cr,1} \le 1$, both systems have a similar linear monotonic suppression.  Interestingly, a break in curvature occurs at $x/x_{cr,1} \sim 0.5$ for both systems.

The generalized Abrikosov-Gorkov (A-G) theory of Fulde and
Maki includes three effects that can break Cooper pairs in a BCS SC in the presence of magnetic moments and magnetic field: (1) spin-polarization of conduction electrons by an applied magnetic field, (2) spin-flip scattering of conduction electrons
by magnetic moments, and (3) spin-polarization of conduction electrons by
the exchange field generated by the applied field or magnetic order mediated by the RKKY
interaction between the magnetic moments. The generalized A-G formula that takes these three effects into account has the following form,
\begin{equation}
 ln(\frac{T_{\rm{c}}}{T_{\rm{c0}}})
 -\Psi\left(\frac{1}{2}
 -0.14(\frac{T_{\rm{c0}}}{T_{\rm{c}}})(\sum_{i=1}^{3}\frac{\alpha_{\rm{i}}}{\alpha_{\rm{cr,i}}})\right)
 -\Psi(\frac{1}{2}) = 0,
 \label{eq:GeneralizedAG}
\end{equation}
where the SC depairing parameters are denoted as $\alpha_{\rm{i}}$'s,
their critical values as $\alpha_{\rm{cr,i}}$'s, and
$T_{\rm{c0}}$ is the superconducting transition temperature of the
parent compound in zero applied magnetic field. Equation~\ref{eq:MultiplePairBreakingEffects} describes the total pair-breaking
effect in the magnetically substituted SC system, which is equivalent to
the pair breaking effect due only to the applied field on the parent
compound:
\begin{equation}
 \sum_{i=1}^{3}\frac{\alpha_{\rm{i}}}{\alpha_{\rm{cr,i}}}
  =   \frac{H_{\rm{c2}}(x,T)}{H_{\rm{c2}}(0,0)}
    + \frac{x}{x_{\rm{cr}}} +  \frac{P}{P_{\rm{cr}}}
  = \frac{H_{\rm{c2}}(0,T)}{H_{\rm{c2}}(0,0)},
 \label{eq:MultiplePairBreakingEffects}
\end{equation}
where $x$ is the concentration of subsituted magnetic ions,
$H_{\rm{c2}}(x,T)$ is the upper critical field for concentration $x$
and temperature $T$, and $P$ is the Pauli polarization term
corresponding to the effect of the magnetic exchange field on the
conduction electrons. Following this argument, the temperature
dependence of the upper critical field for concentration $x$ can be
expressed as,
\begin{equation}
 H_{\rm{c2}}(x,T) \approx   H_{\rm{c2}}(0,T)
                    - H_{\rm{c2}}(0,0)\left(\frac{x}{x_{\rm{cr}}}
                    + \frac{P}{P_{\rm{cr}}}\right).
 \label{eq:Hc2vsxT}
\end{equation}
The Pauli polarization term $P$ = \mbox{$\tau_{\rm{so}}(x \Im
<J_{\rm{z}}>)^2$}, where $\tau_{\rm{so}}$ is the spin-orbit
scattering time, $\Im$ is the s-f exchange interaction parameter, and
$<J_{\rm{z}}>$ is the average value of the total angular momentum along
the direction of the exchange field, which is defined as the z
direction. $<J_{\rm{z}}>$ has a Brillouin-function dependence on $T$
but approaches a constant value as $T \rightarrow 0$\,K, that is
proportional to the magnetic substituent's ``ground-state" magnetic moment. To
simplify the analysis, we focus on the behavior at 0\,K. In
this case Eqns.~\ref{eq:MultiplePairBreakingEffects} and
\ref{eq:Hc2vsxT} are reduced to
\begin{equation}
 H_{\rm{c2}}(x,0) =   H_{\rm{c2}}(0,0)
                    [1-\left( \frac{x}{x_{\rm{cr}}}
                    + \frac{\tau_{\rm{so}}\Im^2 J^2}{P_{\rm{cr}}} \it{x}^{\rm{2}} \right)
                    ].
 \label{eq:Hc2vsxat0K}
\end{equation}
In the low concentration regime, $H_{\rm{c2}}(x,0)$ of
La$_{3-x}$Gd$_x$In has a linear $x$ dependence, but
$H_{\rm{c2}}(x,0)$ of Pr$_{1-x}$Nd$_x$Os$_4$Sb$_{12}$ shows a more
quadratic behavior. The linear dependence of $x$ in the low concentration region of La$_{3-x}$Gd$_x$In indicates that the Gd magnetic moments are in the dilute region and no magnetic ordering has developed so the Pauli polarization contribution is negligible.  This suggests that if
Pr$_{1-x}$Nd$_x$Os$_4$Sb$_{12}$ conforms to BCS SC, magnetic correlations may be significant in the low $x$ region.  According to the
Curie-Weiss analysis, $\Theta$ is negative for $x \lesssim 0.3$, indicating that AFM correlations may be dominant in
this concentration range and that the internal magnetic field
found in the SC state is
associated with AFM order. Thus, the SC in PrOs$_4$Sb$_{12}$ may be influenced by both AFM and FM correlations that may have comparable strengths over certain parts of the phase diagram.
It is possible that the slower than linear suppression of $H_{c2}(0)$ vs $x$ in the range of $x$ from 0 to $x_{cr,2}$ $\approx$ 0.3 is due to the combination of pairbreaking by the applied field and the AFM exchange field, while the
suppression of $H_{c2}(0)$ vs $x$ in the range $x_{cr,2}$ $\approx$ 0.3 to $x_{cr,1}$ $\approx$ 0.6 could be primarily due to the FM exchange field, although the curvature of $H_{c2}(0)$ in this region is not in agreement with the generalized Abrikosov-Gorkov (A-G) theory of
Fulde and Maki.

\section{Summary}
The effect of magnetic moments on the normal and superconducting
states of PrOs$_4$Sb$_{12}$ has been investigated in the
Pr$_{1-x}$Nd$_x$Os$_4$Sb$_{12}$ system. In the normal state, the
feature associated with the HFOP is clearly observed up to $x =
0.5$.  The kink in the lower phase boundary of the HFOP
seems to correlate with a similar feature in $H_{\rm{c2}}(x,0)$.  The HFOP is more robust against substituent concentrations $x$
in Pr$_{1-x}$Nd$_x$Os$_4$Sb$_{12}$ than in the Pr(Os$_{1-x}$Ru$_x$)$_4$Sb$_{12}$ system. The Curie-Weiss
analysis for $T$ between 2\,K and 10\,K suggests that weak FM exists
for $0.3 \lesssim x \lesssim 1$ and AFM correlations are important in the range $x \lesssim 0.3$.

In the SC state, features associated with FM were observed in
the $x = 0.45$ sample. Quadratic and linear dependences of $x$ were found
in $H_{\rm{c2}}(0)$ for $0 \le x \lesssim 0.3$ and $ 0.3 \lesssim x \lesssim 0.6$,
respectively, which indicates that there are two critical concentrations in the
Pr$_{1-x}$Nd$_x$Os$_4$Sb$_{12}$ system, $x_{\rm{cr,1}} \sim 0.6$ and
$x_{\rm{cr,2}} \sim 0.3$. The break in slope for
$H_{\rm{c2}}(x,0)$ may be related to the existence of two bands of
SC electrons in PrOs$_4$Sb$_{12}$, which are affected by Nd
substitution differently. On the other hand, the multiple-pair
breaking effect could also explain the behavior in
$H_{\rm{c2}}(x,0)$ where $x_{\rm{cr,1}}$ and $x_{\rm{cr,2}}$ are associated
with the suppression of SC in the FM and AFM correlation regimes, respectively.

\section{Acknowledgments}
%\begin{acknowledgments}
Research at UCSD was supported by the U.S. Department of Energy under Grant No.
DE-FG02-04ER46105 for single crystal growth and characterization and the National Science Foundation under Grant No. DMR 0802478 for low temperature measurements. Research at California State University, Fresno was supported by Research Corporation under the CCSA Grant No.
7669. Research at Hokkaido University
was supported by a Grant-in-Aid for Young Scientists (B) from MEXT, Japan.
%\end{acknowledgments}

% Create the reference section using BibTeX:
% You should use BibTeX and apsrev.bst for references
% Choosing a journal automatically selects the correct APS
% BibTeX style file (bst file), so only uncomment the line
% beloz if necessary.
%\bibliographystyle{apsrev}
\bibliographystyle{prsty} %this style will cut off the author list
%\bibliography{RefPr1_xNdxO4Sb12Pub1}

%\newpage
% figures of the paper
%   lattice parameter vs Ndx
%   chi_dc vs T (2K-300K)
%   chi_ac vs T (below 2.6 K)
%   T_c & T_FM vs Ndx
%   CoexistenceSCFMNd0_45.eps
%   H_{c2} vs T with various Ndx
%   H-x phase diagram with enlarged Hc2 vs Ndx

%
%\begin{figure}
% \begin{center}
%  \includegraphics[angle=0,width=0.5\textwidth]{}
%   \caption{}
%  \label{}
% \end{center}
%\end{figure}

\end{document}